%% file: main.tex
\documentclass[prd,nofootinbib,notitlepage,10pt,aps]{revtex4-1}
\usepackage{amsmath,amssymb,slashed}
\usepackage[utf8]{inputenc}
\usepackage{graphicx}
\usepackage{hyperref}
\usepackage{bm,bbold}
\usepackage{accents}

\input{commands}
\usepackage{physics}

\usepackage[usenames,dvipsnames,svgnames,table]{xcolor}

\begin{document}
%**************************************************************************************************
\title{Freezeout at constant energy density and spin polarization in heavy-ion collisions}
%**************************************************************************************************
\author{Andrea Palermo}\affiliation{Center for Nuclear Theory, Department of Physics and Astronomy,
Stony Brook University, Stony Brook, New York 11794–3800, USA}
\author{Masoud Shokri}\affiliation{Institut f\"ur Theoretische Physik, 
	Johann Wolfgang Goethe--Universit\"at,
	Max-von-Laue-Str.\ 1, D--60438 Frankfurt am Main, Germany}

\begin{abstract}
Using the local equilibrium density operator, we develop a geometry-informed linear response theory that takes into account the parameterization of the freezeout hypersurface before the gradient expansion is carried out. Assuming local equilibrium on an iso-energy density hypersurface, we study the $\Lambda$ hyperon spin polarization, and compute corrections to the isothermal case due to finite density. We argue that corrections to isothermal freezeout should be small, and that even in heavy ion collisions with energy as low as $\sqrt{s_{NN}}= 11.5$ GeV they constitute at most a $10\%$ effect. They may, however, become relevant for local polarization observables.
\end{abstract}
\maketitle
%***************************************************************************************************

\section{Introduction}
The quark-gluon plasma (QGP) is an almost perfect relativistic fluid created in high-energy heavy-ion collisions. 
In non-central collisions, part of the angular momentum of the QGP---inherited from the initial colliding system---is transferred to the spin of final-state hadrons \cite{Liang:2004ph,Liang:2007ma,Gao:2007bc,Becattini:2007sr}. 
This phenomenon can be probed by various experimental observables, such as the polarization of $\Lambda$ particles and the spin alignment of vector mesons \cite{Becattini:2024uha,Chen:2024afy}.
$\Lambda$ polarization was first observed by the STAR experiment \cite{STAR:2017ckg}, and since then, numerous other measurements have followed \cite{STAR:2019erd,STAR:2023eck,ALICE:2021pzu,HADES:2022enx,STAR:2021beb,STAR:2018gyt,STAR:2017ckg,STAR:2023nvo,ALICE:2019onw}. 
These data span a wide range of collision energies, even showing evidence of spin polarization in pPb collisions \cite{CMS:2025nqr}.

A viable theoretical description of $\Lambda$ polarization is provided by quantum relativistic statistical mechanics through the so-called local equilibrium density operator \cite{Zubarev:1979,vanWeert,Becattini:2019dxo}, which we adopt in this work. 
Alternative approaches to $\Lambda$ polarization include quantum field theory, kinetic theory, and spin hydrodynamics \cite{Liu:2021uhn,Fu:2021pok,Weickgenannt:2020aaf,Wagner:2024fry,Sapna:2025yss,Singh:2024cub}.
Notably, recent developments in semi-classical spin hydrodynamics suggest a rapid equilibration of spin degrees of freedom in moderate to high-energy collisions, supporting the applicability of the local equilibrium framework \cite{Wagner:2024fhf,Chiarini:2024cuv}.

In Refs.\ \cite{Becattini:2021suc,Becattini:2021iol}, the inclusion of shear-induced polarization was shown to restore agreement between local polarization data and theoretical expectations, relying on the additional assumption that the freezeout hypersurface, where the fluid-dynamic description ceases to apply, is isothermal. 
While this assumption is well justified in very high-energy heavy-ion collisions, where chemical potentials are negligible and temperature is the only relevant intensive thermodynamic variable, it becomes questionable at lower energies. 
In such cases, one or more chemical potentials (e.g., the baryon chemical potential) are non-negligible at freezeout, and must be taken into account.

In this work, we develop a linear response framework that incorporates the geometry of the freezeout hypersurface. We focus on the case where freezeout occurs at constant energy density, which is common practice. Nonetheless, our method is readily generalizable to other freezeout conditions. The constant-energy-density assumption introduces thermodynamic constraints that relate variations of temperature and chemical potentials along the freezeout hypersurface. These relations can be exploited to compute gradient corrections to the mean spin vector of particles. We derive modified expressions for the $\Lambda$ polarization that reflect these geometric-thermodynamic constraints, and show that they reduce to the isothermal freezeout formulae of Ref.\ \cite{Becattini:2021iol} when the equation of state is independent of chemical potential. We assess the magnitude of these geometric corrections using two different equations of state and find that they are modest---at most $\sim 10\%$---in low-energy collisions. 
A full hydrodynamic implementation will be pursued in future work.

This paper is organized as follows. 
After a brief review of the spin polarization calculation using the local equilibrium density operation in Sec.\ \ref{sec:rev-spin-calc}, we introduce our geometric approach in Sec.\ \ref{sec:geometric-aprox} and apply it to compute the spin vector of $\Lambda$ particles within linear response theory.
In Sec.\ \ref{sec:evaluation}, we evaluate the geometric corrections coefficient for two different equations of state, as a function of temperature and chemical potential.
Finally, Sec.\ \ref{sec:conclusions} summarizes our findings. 
Supporting material is provided in the Appendix \ref{app:coefficient}. 
%***************************************************************************************************

%%%%%%%%%%%%%%%%%%%%%
\paragraph*{Notations and conventions}
In this paper, we adopt natural units setting $\hbar = c = k_B = 1$. The Minkowski metric tensor is taken as $g = \mathrm{diag}(1, -1, -1, -1)$, and we use the convention $\epsilon^{0123} = 1$ for the Levi-Civita symbol.
We employ relativistic index notation, with repeated indices implicitly summed.
Dirac matrices are denoted by $\gamma^\mu$.
Operators in Hilbert space are indicated with a wide upper hat, e.g., $\widehat{H}$, except for the Dirac field operator $\Psi$ and its adjoint $\bar{\Psi} = \Psi^\dagger \gamma^0$.
Traces over Hilbert space are written with a capital letter, $\Tr$, while traces over finite-dimensional matrix spaces use lowercase, $\tr$.
Expectation values are denoted with angle brackets, $\langle \widehat{O} \rangle = \Tr(\wrho\, \widehat{O})$, and normal ordering is indicated by colons, $\langle :\widehat{O}: \rangle$.

The common hydrodynamic quantities are denoted as follows: $u_\mu$ is the fluid four-velocity normalized as $u^\mu u_\mu = 1$, $e$ is the energy density, $P$ is the pressure, $T$ is the temperature, and $\mu$ is the chemical potential.
$\beta^\mu=u^\mu/T$ is the four-temperature vector with $T=1/\sqrt{\beta\cdot\beta}$, and $\zeta=\mu/T$.
We use the standard symmetrization and antisymmetrization notations, $A_{(\mu\nu)} \equiv \tfrac{1}{2}\left(A_{\mu\nu} +A_{\nu\mu}\right)$ and
$A_{[\mu\nu]} \equiv \tfrac{1}{2}\left(A_{\mu\nu} -A_{\nu\mu}\right)$, respectively.
We define the kinematic vorticity $\omega_{\mu\nu}\equiv\partial_{[\nu}u_{\mu]}$, the kinematic shear $\Xi_{\mu\nu}\equiv \partial_{(\nu}u_{\mu)}$, the thermal vorticity $\varpi_{\mu\nu}\equiv\partial_{[\nu}\beta_{\mu]}$ and thermal shear $\xi_{\mu\nu}\equiv \partial_{(\nu}\beta_{\mu)}$.
%***************************************************************************************************

\section{Brief review of polarization at local equilibrium in linear response theory}\label{sec:rev-spin-calc}
In this section, we summarize the main steps in deriving the mean spin vector as carried out in Refs.\ \cite{Becattini:2021suc,Becattini:2021iol}. 
The mean spin vector for a spin-$1/2$ field is given by the following formula (c.f.\ Eq.\ (71) of Ref.\ \cite{Becattini:2020sww}):
\begin{equation}\label{eq:mean-spin-vector}
    S^\mu(p)=\frac{\int_{\Sigma_D} \dd{\Sigma}\cdot p \tr\left[\gamma^\mu\gamma_5 W_+(x,p)\right]}{\int_{\Sigma_D} \dd{\Sigma}\cdot p \tr\left[W_+(x,p)\right]}\,,
\end{equation}
where $\dd{\Sigma}_\mu$ is the surface element on the freezeout hypersurface $\Sigma_D$.
The function $W_+$ represents the particle component of the Wigner function, defined as \cite{DeGroot:1980dk}
\begin{equation}
    W_+(x,k)=-\frac{\theta(k^2)\theta(k^0)}{(2\pi)^4}\int \di^4 y e^{-ik\cdot y}\langle:\Psi(x-y/2)\bar{\Psi}(x+y/2):\rangle\,.
\end{equation}
It corresponds to the expectation value of the Wigner operator $\widehat{W}_+(x,k)$,
\begin{equation}\label{eq:wplus}
    W_+(x,k)=\langle\widehat{W}_+(x,k)\rangle=\Tr\left(\wrho \,\widehat{W}_+(x,k)\right)\,,
\end{equation}
where $\wrho$ is the density operator. 
If the density operator is known, the Wigner function is then unambiguously determined. However, obtaining it is challenging in practice.

In \textit{global equilibrium}, the exact calculation of the Wigner function is possible using the analytic distillation technique   \cite{Becattini:2020qol,Palermo:2021hlf,Palermo:2023cup}; global equilibrium in a relativistic quantum system is achieved when the density operator takes the form
\begin{align}\label{eq:geq-rho}
\wrho = \frac{1}{Z} \exp(-b\cdot \wP +\frac{1}{2}\varpi:\wJ+\zeta \wQ),    
\end{align}
where $\zeta$, $b_\mu$ and $\varpi_{\mu\nu}$ are constant scalar, four-vector and antisymmetric tensor Lagrange multipliers, respectively, and $\wP_\mu$, $\wJ_{\mu\nu}$, and $\wQ$ are the four-momentum, boost-angular momentum and scalar charge (e.g., electric charge) operators.
In Eq.\ \eqref{eq:geq-rho}, $Z$ is the partition function given by $Z=\Tr[\exp(-b\cdot \wP +\frac{1}{2}\varpi:\wJ+\zeta \wQ)]$, such that $\Tr(\wrho)=1$. 
A nonvanishing thermal vorticity tensor $\varpi_{\mu\nu}$ causes the four-temperature vector $\beta_\mu$ to be space-time dependent, given by
\begin{align}
    \beta_\mu = b_\mu +\varpi_{\mu\nu} x^\nu\;,
\end{align}
which satisfies the Killing condition $\partial_{(\mu}\beta_{\nu)}=0$.
The scalar temperature is defined as $T=1/\sqrt{\beta^2}$, whence the four-velocity of the fluid is $u^\mu=T\beta^\mu$.
The spacetime dependence of $\beta_\mu$ implies that $u^\mu$ corresponds to a rigidly rotating and/or accelerating flow \cite{Shokri:2023rpp}. 
The scalar $\zeta$ is the ratio between the chemical potential $\mu$ and the scalar temperature, i.e., $\zeta=\mu/T$.

However, in heavy-ion collisions, the assumption of global equilibrium is unrealistic. 
A better description of heavy-ion collisions can be achieved by assuming that the final hadrons, such as $\Lambda$ hyperons, are produced in a state of \textit{local thermodynamic equilibrium}, 
where each fluid cell can be approximated as an infinitesimally small thermodynamic system.
Consequently, applying the maximum entropy principle locally to each fluid cell yields the following expression for the density operator~\cite{Zubarev:1979,vanWeert,Becattini:2019dxo}:
\begin{equation}\label{eq:rho-le}
    \wrho = \frac{1}{Z}\exp\left\{-\int _{\Sigma_D}\di \Sigma_{\mu}(y) \left[\wT^{\mu\nu}(y)\beta_\nu(y)-\wj^\mu(y) \zeta(y)\right]\right\}\,,
\end{equation}
where $\wT^{\mu\nu}$ and $\wj^\mu$ are the energy-momentum and charge density current operators, respectively.
Regarding the Lagrange multipliers, $\zeta$ is not a constant, but a locally defined scalar field, and $\beta_\mu$ is no longer a Killing vector.
The density operator \eqref{eq:rho-le} involves an integration over the freezeout hypersurface $\Sigma_D$. 
Consequently, when computing the mean spin with equation \eqref{eq:mean-spin-vector}, the integral on this surface appears twice, a point that will become important later.
Note that we adopt the Belinfante pseudogauge, where the energy-momentum tensor is symmetric and the spin tensor vanishes.

To calculate the mean spin vector using the local equilibrium density operator, we first insert Eq.\ \eqref{eq:rho-le} into Eq.\ \eqref{eq:wplus} to obtain,
\begin{equation}
    W^{(LE)}_+(x,k) = \frac{1}{Z}\Tr\left\{\widehat{W}_+(x,k)\,\exp\left[-\int_{\Sigma_{D}}\di\Sigma_\mu(y) \left(\wT^{\mu\nu}(y)\beta_\nu(y) - \wj^\mu(y) \zeta(y)\right)\right]\right\}\,.
\end{equation}
This expression can be recast as
\begin{equation}\label{eq:expansion-with-variations}
    W^{(LE)}_+(x,k) = \frac{1}{Z}\Tr\left\{\widehat{W}_+(x,k)\,\exp\left[-\beta(x)\cdot \wP+\zeta(x) \wQ -\int_{\Sigma_{D}}\di\Sigma_\mu(y) \wT^{\mu\nu}(y)\delta\beta_\nu(x,y) - \wj^\mu(y) \delta\zeta(x,y)\right]\right\}\,,
\end{equation}
where  $\delta\beta$ and $\delta\zeta$ represent the \emph{exact} variations of the thermodynamic field between the points $x$ and $y$,
\begin{equation}\label{eq:variations}
    \delta\beta_\nu(x,y) = \beta(y)-\beta(x)\,, \qquad \qquad \delta \zeta(x,y) = \zeta(y)-\zeta(x)\,.
\end{equation}
In Eq.\ \eqref{eq:expansion-with-variations}, we used the definitions of the total four-momentum and charge operators
$$
\wP^\mu = \int _{\Sigma_D}\di \Sigma_{\mu} \wT^{\mu\nu}\,,\qquad \wQ = \int _{\Sigma_D}\di \Sigma_{\mu}(y) \wj^\mu\,.$$  

We now assume that the terms involving the variations of the fields in Eq.\ \eqref{eq:variations} contribute to the Wigner function as small corrections compared to the leading terms involving the fields themselves.
Namely, we expand the exponential in Eq.\ \eqref{eq:expansion-with-variations} within the framework of linear response theory; the details of this procedure can be found, for example, in Refs.~\cite{Becattini:2019dxo, Becattini:2021suc}. 
With this approach, the local equilibrium Wigner function \eqref{eq:expansion-with-variations} can be approximated as:
\begin{align}\label{eq:lr-wigner-function}
     W_{LE}(x,k) &\simeq \langle \widehat{W}(x,k)\rangle_{\beta(x),\zeta(x)}
     -\int_0^1 \di z\int_{\Sigma_{D}}\di\Sigma_\mu(y)\left[\delta\beta_\nu(x,y)\langle \widehat{W}(x,k)\wT^{\mu\nu}(y+iz\beta(x))\rangle^c_{\beta(x),\zeta(x)}\right.\\\nonumber
     -&\left. \delta\zeta(x,y)\langle \widehat{W}(x,k)\wj^\mu(y+iz\beta(x))\rangle^c_{\beta(x),\zeta(x)}\right]\,.
\end{align}
Here,
\begin{equation}\label{eq:global-exp-at-x}
    \expval{\widehat{W}(x,k)\widehat{X}}_{\beta(x),\zeta(x)} \equiv \frac{1}{Z}\Tr\left[\widehat{W}(x,k)\widehat{X}\,\exp(-\beta(x)\cdot \wP+\zeta(x)\wQ)  \right]\,,
\end{equation}
where $\widehat{X} \in \{ \widehat{1}, \wT^{\mu\nu}(y), \wj^{\mu}(y)\}\,$, and the superscript $c$ denotes the connected two-point function: $$\langle \widehat{O}\widehat{X}\rangle^c = \langle \widehat{O}\widehat{X}\rangle -\langle \widehat{O}\rangle\langle\widehat{X}\rangle\;.$$

This procedure can be used to calculate the leading-order corrections (in the field variations eq. \eqref{eq:variations}) to the Wigner function, and expectation values can be computed therefrom. These steps are general and hold for the Wigner function of any field.
From now on, we will focus on the calculation of the mean spin of Dirac fermions. 

Noting that the leading term in Eq. \eqref{eq:lr-wigner-function} does not contribute to the numerator of Eq.\ \eqref{eq:mean-spin-vector}, we insert Eq.\ \eqref{eq:lr-wigner-function} into Eq.\ \eqref{eq:mean-spin-vector} to obtain the mean spin vector at local equilibrium:
\begin{align}\label{eq:general-linear-response-spin}
    S^\mu(p)=&-\frac{\int_{\Sigma_D} \di\Sigma(x)\cdot p\int_0^1 \di z 
\int_{\Sigma_{D}}\di\Sigma_\rho(y)
\delta\beta_\nu(x,y)\tr\left(\gamma^\mu\gamma_5\langle \widehat{W}(x,k)\wT^{\rho\nu}(y+iz\beta(x))\rangle^c_{\beta(x),\zeta(x)}\right)}{\int_{\Sigma_D}\di\Sigma\cdot p \tr\left(\langle \widehat{W}(x)\rangle_{\beta(x),\zeta(x)}\right)}\,\nonumber\\
    &+\frac{\int_{\Sigma_D} \di\Sigma(x)\cdot p\int_0^1 \di z 
\int_{\Sigma_{D}}\di\Sigma_\rho(y)\delta\zeta(x,y)\tr\left(\gamma^\mu\gamma_5\langle \widehat{W}(x,k)\wj^\rho(y+iz\beta(x))\rangle_{\beta(x),\zeta(x)}^c\right)
    }{\int_{\Sigma_D}\di\Sigma\cdot p \tr\left(\langle \widehat{W}(x)\rangle_{\beta(x),\zeta(x)}\right)}\,.
\end{align}

Up to now, the variations $\delta\beta_\nu$ and $\delta \zeta$ are exact variations, but it is not possible to proceed further without approximating them in some way. For instance, as it is usually done, we can Taylor-expand the components of $\beta_\mu(y)$ and $\zeta(y)$ around $x$, which is the point where the Wigner function is evaluated:
\begin{align*}
    \delta\beta_\nu\approx (y-x)^\alpha\partial_\alpha\beta_\nu\, ,\qquad \qquad
    \delta \zeta \approx (y-x)^\alpha\partial_\alpha \zeta\,.
\end{align*}
This is justified, because the two point functions $\langle\widehat{W}(x,k)\widehat{X}(y+i\beta(x))\rangle^c_{\beta(x),\zeta(x)}$ vanish for large space separations $(x-y)$, and in the hydrodynamic regime $\delta\beta$ and $\delta\zeta$ are small on the length scale where the two point function is non-zero.
Upon using this approximation, one finds \cite{Becattini:2021suc,Buzzegoli:2022qrr}\footnote{In Ref.\ \cite{Buzzegoli:2022qrr}, a typo in Eq.\ 44 was found, and the factor 2 in the denominator of the spin hall effect should have been in the numerator.}
\begin{equation}\label{eq:total-spin}
    S^\mu(p) = S_A^\mu(p)+S_S^\mu(p)+S_\zeta^\mu(p)\,,
\end{equation}
where the antisymmetric, symmetric, and spin-Hall contributions, are labeled as $A$, $S$, and $\zeta$ respectively.
They read
\begin{subequations}\label{eq:taylor-expansion-pol}
    \begin{align}
    &S_A^\mu(p) =-\frac{1}{8m\,\mcN}\epsilon^{\mu\nu\rho\sigma}p_\sigma{\int_{\Sigma_D} \di \Sigma \cdot p\; \varpi_{\nu\rho}n_F(p)(1-n_F(p))}\,,\\
    &S_S^\mu(p) =-\frac{1}{4m\,\mcN}\epsilon^{\mu\nu\rho\sigma}p_\sigma\frac{p^\lambda}{p^0}\hat{t}_\nu\int_{\Sigma_D} \di \Sigma \cdot p\; \xi_{\rho\lambda}n_F(p)(1-n_F(p))\,,\\
    &S^\mu_\zeta(p) = +\frac{1}{4m\,\mcN}\epsilon^{\mu\nu\rho\sigma}p_\sigma\frac{\hat{t}_\nu }{p^0}\int_{\Sigma_D} \di \Sigma \cdot p\; \partial_\rho \zeta n_F(p)(1-n_F(p))\,.
    \end{align}
\end{subequations}
Here, $n_F = \left[e^{(p\cdot u-\mu)/T}+1\right]^{-1}$ is the Fermi-Dirac distribution, $\hat{t}_\nu = \left(1,\vb{0}\right)$ in Cartesian coordinates, $p^0 = p\cdot \hat{t}$, and we have defined
\begin{equation}
    \mcN \equiv \int_{\Sigma_D}\di\Sigma \cdot p\, n_F(p)\,.
\end{equation}

While general, the derivation of Eq.\ \eqref{eq:taylor-expansion-pol} may not be the most accurate in specific cases.
Indeed, in Ref.\ \cite{Becattini:2021iol} it was observed that a more precise calculation of the mean spin vector in high-energy collisions can be achieved by noting that the freezeout hypersurface is approximately isothermal, i.e., the temperature at all points on $\Sigma_D$ satisfies $T(x)=T_D\,$, where $T_D$ is constant.

Consequently, using the definitions of the Lagrange multipliers $\beta_\mu$ and $\zeta$, we find that the exact variations are 
\begin{align*}
    &\delta\beta_\nu = \beta_\nu(y)-\beta_\nu(x) = \frac{1}{T_D}(u_\nu(y)-u_\nu(x)) = \frac{\delta u_\nu}{T_D}\,,\\
    &\delta\zeta = \zeta(y)-\zeta(x) = \frac{1}{T_D}(\mu(y)-\mu(x)) = \frac{\delta \mu}{T_D}\,,
\end{align*}
where $\delta u^\nu$ and $\delta \mu$ are the exact variations of four-velocity and chemical potential.
Therefore, to proceed further, we approximate only the latter variations using a first-order Taylor expansion, but keep the temperature constant:
\begin{align}\label{eq:iso-t-vars}
    &\delta\beta_\nu \approx (y-x)^\alpha\frac{\partial_\alpha u_\nu}{T_D}\,,
    \qquad \qquad
    \delta\zeta \approx (y-x)^\alpha\frac{\partial_\alpha \mu}{T_D}\,.
\end{align}
Substituting these expressions into Eq.\ \eqref{eq:general-linear-response-spin} yields the results of Ref.\ \cite{Becattini:2021iol}, expressed in terms of kinematic vorticity $\omega$ and kinematic shear $\Xi$.
For the chemical potential sector, the result agrees with those of Refs.\ \cite{Buzzegoli:2022qrr,Ivanov:2022geb} after substituting $\partial\zeta$ with $\partial \mu /T_D$:
\begin{subequations}\label{eq:isothermal-pol}
    \begin{align}
    &S_A^\mu(p) =-\frac{1}{8m\, \mcN T_D}\epsilon^{\mu\nu\rho\sigma}p_\sigma\int_{\Sigma_D} \di \Sigma \cdot p\; \omega_{\nu\rho}n_F(p)(1-n_F(p))\,,\\
    &S_S^\mu(p) =-\frac{1}{4m\,\mcN T_D}\epsilon^{\mu\nu\rho\sigma}p_\sigma\frac{p^\lambda}{p^0}\hat{t}_\nu\int_{\Sigma_D} \di \Sigma \cdot p\; \Xi_{\rho\lambda}n_F(p)(1-n_F(p))\,,\\
    &S^\mu_\zeta(p) = +\frac{1}{4m\,\mcN T_D}\epsilon^{\mu\nu\rho\sigma}p_\sigma \frac{\hat{t}_\nu}{p^0}\int_{\Sigma_D} \di \Sigma \cdot p\; \partial_\rho \mu \;n_F(p)(1-n_F(p))\,.
    \end{align}
\end{subequations} 
We emphasize that temperature gradients have not been neglected during the calculations; substitutions \eqref{eq:iso-t-vars} follow solely from assuming that the freezeout surface is isothermal and from the fact that both points $y$ and $x$ lie on this surface. 
While the components of the temperature gradient tangent to the isothermal surface vanish, the normal component does not; it defines the vector normal to the surface via $n_\mu \propto \partial_\mu T$.

Equation \eqref{eq:isothermal-pol} illustrates how adjusting the variations of the Lagrange multipliers based on geometric considerations refines the final polarization formula. 
However, the assumption of an isothermal freezeout hypersurface applies only to high-energy heavy-ion collisions, and it is not a reasonable approximation at more moderate collision energies. 
In practice, the freezeout hypersurface is often defined as the set of fluid cells with a constant energy density. 
In high-energy collisions, the chemical potential at freezeout is negligible, making the iso-energy density and isothermal hypersurfaces effectively equivalent to a very good precision. 
This approximation breaks down when the chemical potential becomes comparable to the temperature, which occurs at collision energies below $\sqrt{s_{NN}} \lesssim 39$ GeV (see, e.g., ref \cite{STAR:2017sal}).
Consequently, Eq.\ \eqref{eq:isothermal-pol} is no longer a reliable approximation for the spin vector of $\Lambda$ particles at those energies and must be reconsidered.
In the next section, we introduce an approach to systematically address this problem.

\section{Geometric Approximations on the Freezeout Hypersurface}\label{sec:geometric-aprox}

In this section, we develop a scheme to more accurately account for the variations of thermodynamic quantities on the freezeout hypersurface $\Sigma_D$,  which is defined as the iso-energy density level surface:
\begin{equation}\label{eq:level-surface}
    e(x)=e_0\,,
\end{equation}
where $e_0$ is a constant representing the critical energy density.
The gradient of the function $e(x)$ determines the normal vector $n_\mu$ to $\Sigma_D$.
One may introduce a set of coordinates $\lambda^a$ on $\Sigma_D$, with $a=1,2,3$, such that the tangent vectors to $\Sigma_D$ are given by
\begin{equation}
    \mathrm{e}^{\mu}_a = \pdv{x^\mu}{\lambda^a}\;,
\end{equation}
satisfying 
\begin{equation}\label{eq:tangent-diff}
    \mathrm{e}^{\mu}_a \partial_\mu e = 0\;.
\end{equation}
The energy density is given, via the equation of state, as a function of $T$ and $\mu$ (assuming a single conserved charge).
Thus, the variation of the energy density can be expressed as
\begin{equation}\label{eq:differential-energy}
    \delta e=\left.\frac{\partial e}{\partial T}\right|_\mu \delta T +\left.\frac{\partial e}{\partial \mu}\right|_T \delta \mu\,,
\end{equation}
where the subscript indicates the thermodynamic variable held constant in each partial derivative.
On the hypersurface $\Sigma_D$, Eq. \eqref{eq:tangent-diff} implies $\delta e = 0$, and therefore
\begin{equation}\label{eq:hypersurf-variations}
    \left.\frac{\partial e}{\partial T}\right|_\mu \delta T =-\left.\frac{\partial e}{\partial \mu}\right|_T \delta \mu\,.
\end{equation}

This equation, which relates tangent variations of temperature and chemical potential, holds strictly on the hypersurface.
In principle, if the hypersurface coordinates $\lambda^a$ are known, one can Taylor-expand the quantities in the integrand in Eq.\ \eqref{eq:rho-le} in $\lambda^a$, thereby treating only the \emph{hypersurface variations}.
However, this is not feasible in hydrodynamic simulations, where $\lambda^a$ and thus the tangent derivatives of $T$ and $\mu$ are not explicitly available.
Therefore, in practice, one resorts to using \textit{spacetime variations}, i.e., variations computing Taylor expanding in with respect to the Minkowski coordinates $x^\mu$ rather than the hypersurface coordinates $\lambda^a$. 
This necessarily introduces an approximation, as the spacetime and hypersurface variations can differ substantially.

To see how using Eq.\ \eqref{eq:hypersurf-variations} mitigates this problem, we return to the case of high-energy heavy-ion collisions, where the chemical potential is negligible.
Suppose the simulation employs an equation of state depending on both temperature and chemical potential, and the freezeout hypersurface is identified with equation \eqref{eq:level-surface}. 
Since we are considering a very high-energy heavy-ion collision, the dependence of the energy density on chemical potential along the freezeout hypersurface is negligible.
As a result, Eq.\ \eqref{eq:hypersurf-variations} yields $\delta T\simeq 0$. 
This shows that an error would have been introduced had we approximated the hypersurface variation with the spacetime variation without employing Eq.\ \eqref{eq:hypersurf-variations}.
The error is of order $\order{\partial T}$, and since the polarization vector is itself of order $\order{\partial}$, neglecting Eq.\ \eqref{eq:hypersurf-variations} introduces a significant relative error.
On the other hand, the hypersurface variation of the chemical potential $\delta\mu$ remains unconstrained, so the best approximation at our disposal, and the most natural one, is to replace it with its spacetime variation: $\delta\mu \simeq  (y - x)^\alpha \partial_\alpha \mu$.

This example illustrates that, although hypersurface and spacetime variations are not generally equivalent, the hypersurface variation of either temperature or chemical potential can still be approximated by its spacetime variation.
The choice of which hypersurface variation to approximate with the corresponding spacetime variation should be guided by the geometry of the hypersurface and the underlying physical conditions.

\subsection{Spacetime versus hypersurface variations}

We now return to the general case, where the chemical potential is not negligible.
At an arbitrary point on the hypersurface, where neither $\left.\partial e/\partial T\right|_\mu$ nor $\left.\partial e/\partial \mu\right|_T$ vanish, Eq.\ \eqref{eq:hypersurf-variations} can be rewritten as
\begin{equation}\label{eq:variations-on-the-hyper}
    \delta T =-\mathcal{C}\delta\mu\,.
\end{equation}
Here, we have defined the following thermodynamic coefficient:
\begin{equation}\label{eq:def-C}
    \mathcal{C}=\frac{\left.\partial e/\partial \mu\right|_{T}}{\left.\partial e /\partial T\right|_\mu}\,,
\end{equation}
which can also be expressed in terms of other thermodynamic coefficients, as shown in Appendix \ref{app:coefficient}\,.

We point out that, in our previous illustrative example of high-energy collisions, $\abs{\mathcal{C}} \ll 1$, and consequently $\abs{\delta T} \ll \abs{\delta \mu}$ on $\Sigma_D$.
This suggests that the magnitude of $\mathcal{C}$ determines which variation is larger on the hypersurface.
It can therefore serve as a guiding criterion: the larger of the two variations is approximated by its spacetime counterpart via a Taylor expansion, while the smaller is determined using Eq.\ \eqref{eq:variations-on-the-hyper}.

To clarify this criterion, let us consider the case $|\mathcal{C}|<1$, which implies that on $\Sigma_D$
\begin{equation}
    |\delta T|<|\delta \mu|\,.
    \label{eq:inequality-sup-1}
\end{equation}
Following our guiding principle, we approximate $\delta \mu$ by its spacetime variation and write 
\begin{equation}
    \delta T = -\mathcal{C} (y-x)^\alpha\partial_\alpha \mu \,.\label{eq:c-less-1}
\end{equation}
In the hydrodynamic regime, where $\partial T$ and $\partial \mu$ are of comparable magnitude, this expression ensures that the approximation error for $\delta T$ is suppressed by $\mathcal{C}$: \footnote{Note that in the case of an isothermal freezeout hypersurface, where $\mathcal{C}=0$, we recover $\delta T=0$.}
\begin{equation*}
    \abs{\delta T} = \abs{\mathcal{C} (y-x)^\alpha \partial_\alpha \mu} < \abs{(y-x)^\alpha\partial_\alpha \mu} \sim \abs{(y-x)^\alpha\partial_\alpha T }\,.
\end{equation*}
In contrast, in the case $|\mathcal{C}|>1$, our guiding principle implies expressing the variation of the chemical potential in terms of the spacetime variation of the temperature:
\begin{equation}\label{eq:c-great-1}
    \delta \mu = -\frac{1}{\mathcal{C}}(y-x)^\alpha\partial_\alpha T \,.
\end{equation}
This again suppresses the approximation error in $\delta \mu$, improving the overall accuracy of the polarization formula. \footnote{The limiting case of this formula $\mathcal{C}\to\infty$ corresponds to an iso-chemical-potential freezeout.}

Equations \eqref{eq:c-less-1} and \eqref{eq:c-great-1} can now be used to approximate the variations in Eqs.\ \eqref{eq:variations} and \eqref{eq:general-linear-response-spin} as
\begin{subequations}\label{eq:variations2}
\begin{align}
&\delta\beta_\mu\sim
\frac{\delta u_\nu}{T}-u_\nu \frac{\delta T}{T^2}
\sim 
\begin{cases}
  (y-x)^\alpha\,\partial_\alpha \beta_\mu(x)\,,&\qq{if }|\mathcal{C}| > 1, \\
  {T}^{-1}(y-x)^\alpha\,\left[\partial_\alpha u_\mu(x) +\mathcal{C}\beta_\mu  \partial_\alpha \mu\right] \,, &\qq{if }|\mathcal{C}| < 1\,,
\end{cases}\label{eq:beta-variation}\\
&\delta\zeta
\sim\frac{\delta\mu}{T}-\mu\frac{\delta T}{T^2}
\sim
    \begin{cases}
        -{T}^{-1}(y-x)^{\alpha}\,\left(\frac{1}{\mathcal{C}}+\zeta\right)\partial_\alpha T\,, &\qq{if } |\mathcal{C}|>1\,,\\
        +{T}^{-1} (y-x)^\alpha\,\left(1+\zeta \mathcal{C}\right)\partial_\alpha \mu \,, &\qq{if } |\mathcal{C}|<1\,.
    \end{cases} \label{eq:zeta-variation}
\end{align}
\end{subequations}

We also emphasize that the hypersurface variation of the four-velocity is directly expressed via its spacetime variations. 
This is because the four-velocity does not play a role in determining freezeout, and thus it is unconstrained by the definition of $\Sigma_D$. 

Inserting Eq.\ \eqref{eq:variations2} into Eq.\ \eqref{eq:general-linear-response-spin}, we obtain the contributions to the mean spin arising from the antisymmetric and symmetric gradients of the four-temperature, as well as from the spin-Hall effect. 
In the $\mathcal{C}>1$ regime, this yields
\begin{subequations}\label{eq:pol-low-energy}
\begin{align}
 S^\mu_A&=- \epsilon^{\mu\rho\sigma\tau} \frac{p_\tau}{8m} \frac{1}{  \mcN}
  \int_{\Sigma_D} \di \Sigma \cdot p \; n_F (1 -n_F) 
   \varpi_{\rho\sigma}\,,\\
    S^\mu_S&= - \epsilon^{\mu\rho\sigma\tau} \frac{p_\tau\, \hat t_\rho p^\lambda}{4m p^0} \frac{1}{ \mcN}
  \int_{\Sigma_D} \di \Sigma \cdot p \; n_F (1 -n_F) 
   \xi_{\lambda\sigma}\,,
  \\
    S^\mu_\zeta&=- \epsilon^{\mu\rho\sigma\tau} \frac{p_\tau\, \hat t_\rho }{4m p^0} \frac{1}{  \mcN}
  \int_{\Sigma_D} \di \Sigma \cdot p \; n_F (1 -n_F) 
  \frac{\partial_\sigma T}{T}\left(\frac{1}{\mathcal{C}}+\zeta\right)\,.
\end{align}    
\end{subequations}
The first two terms, $S^\mu_A$ and $S^\mu_S$, correspond to polarization induced by thermal vorticity and thermal shear, in agreement with Ref.\ \cite{Becattini:2021suc}.
The spin Hall contribution $S^\mu_\zeta$, on the other hand, is here expressed in terms of temperature gradients---an outcome of the geometric prescription adopted in the $\mathcal{C}>1$ regime.

In the $\abs{\mathcal{C}} < 1$ regime, on the other hand, by plugging Eq.\ \eqref{eq:variations2} into Eq.\ \eqref{eq:general-linear-response-spin}, we obtain
\begin{subequations}\label{eq:pol-high-energy}
\begin{align}
 S^\mu_A&=- \epsilon^{\mu\rho\sigma\tau} \frac{p_\tau}{8m} \frac{1}{\mcN}
  \int_{\Sigma_D} \di \Sigma \cdot p \; n_F (1 -n_F) 
  \frac{1}{T}\left[ \omega_{\rho\sigma} + \mathcal{C}\beta_{[\rho}\partial_{\sigma]} \mu \right]\,,\\
    S^\mu_S&= - \epsilon^{\mu\rho\sigma\tau} \frac{p_\tau\, \hat t_\rho p^\lambda}{4m p^0} \frac{1}{\mcN}
  \int_{\Sigma_D} \di \Sigma \cdot p \; n_F (1 -n_F) 
  \frac{1}{T}\left[ \Xi_{\lambda\sigma} + \mathcal{C}\beta_{(\lambda}\partial_{\sigma)} \mu \right]\,,
  \\
    S^\mu_\zeta&=+ \epsilon^{\mu\rho\sigma\tau} \frac{p_\tau\, \hat t_\rho }{4m p^0} \frac{1}{\mcN}
  \int_{\Sigma_D} \di \Sigma \cdot p \; n_F (1 -n_F) 
  \frac{\partial_\sigma \mu}{T} \left(1+\zeta \mathcal{C}\right)\,.
\end{align}
\end{subequations}
Notably, all contributions to the mean spin vector now depend explicitly on $\mathcal{C}$, and in the limit $\mathcal{C} \to 0$, these expressions reduce to those valid for an isothermal freezeout hypersurface, as presented in Eq.\ \eqref{eq:isothermal-pol}.

\subsection{Quantifying the approximation error}
Which of the two sets of expressions, Eq.\ \eqref{eq:pol-low-energy} or Eq.\ \eqref{eq:pol-high-energy}, should be used to compute the total spin vector in Eq.\ \eqref{eq:total-spin} depends on the value of $\mathcal{C}$. This, in turn, implies that the appropriate polarization formula in heavy-ion collisions must be selected based on the specific characteristics of the collision---such as the center-of-mass energy, collision centrality, and other relevant physical conditions.

To quantify the error arising from an inappropriate choice of approximation, we evaluate the difference between the $\mathcal{C}>1$ and $\mathcal{C}<1$ expressions while keeping the freezeout hypersurface fixed. 
Denoting this difference as $\Delta S = S^{\mathcal{C}>1} - S^{\mathcal{C}<1}$, we find:
\begin{subequations}\label{eq:pre-implicit-dS}
    \begin{align}
        \Delta S_A &=+ \epsilon^{\mu\rho\sigma\tau} \frac{p_\tau}{8m} \frac{1}{\mcN}
  \int_{\Sigma_D} \di \Sigma \cdot p \; n_F (1 -n_F) 
   \frac{1}{T^2}u_{[\rho}(\partial_{\sigma]} T+\C \partial_{\sigma]}\mu)\, ,\\
        \Delta S_S &= \, + \epsilon^{\mu\rho\sigma\tau} \frac{p_\tau\, \hat t_\rho p^\lambda}{4m p^0} \frac{1}{\mcN}
  \int_{\Sigma_D} \di \Sigma \cdot p \; n_F (1 -n_F) 
   \frac{1}{T^2}u_{(\lambda}(\partial_{\sigma)} T+\C \partial_{\sigma)}\mu)\,,\\
        \Delta S_\zeta &= - \epsilon^{\mu\rho\sigma\tau} \frac{p_\tau\, \hat t_\rho }{4m p^0} \frac{1}{\mcN}
  \int_{\Sigma_D} \di \Sigma \cdot p \; n_F (1 -n_F) \frac{1}{T}
  \left[\partial_\sigma T\left(\frac{1}{\mathcal{C}}+\zeta\right)+\partial_\sigma \mu\left(1+\zeta\C\right)\right]\,.
    \end{align}
\end{subequations}
We now assume that the energy density is a continuous and differentiable function of $T$ and $\mu$ in some neighborhood $\mathcal{S}_0$ of the point $Q_0=(T_0,\mu_0)$, where $\partial e/\partial \mu \neq  0$. \footnote{If this derivative vanishes, one can instead express $T$ as a function of $e$ and $\mu$ and follow a similar argument.}
Then, by the implicit function theorem, we can express the Cartesian gradient of $\mu$ in terms of those of $e$ and $T$ in $\mathcal{S}_0$ and write Eq.\ \eqref{eq:pre-implicit-dS} as
\begin{subequations}\label{eq:post-implicit-dS}
    \begin{align}
        \Delta S_A &=+ \epsilon^{\mu\rho\sigma\tau} \frac{p_\tau}{8m} \frac{1}{\mcN}
  \int_{\Sigma_D} \di \Sigma \cdot p \; n_F (1 -n_F) 
   \frac{1}{T^2}u_{[\rho}(\partial_{\sigma]} T+\C \left.\frac{\partial \mu}{\partial e}\right|_T \partial_{\sigma]}e+\C \left.\frac{\partial \mu}{\partial T}\right|_e \partial_{\sigma]}T)\, ,\\
        \Delta S_S &= \, + \epsilon^{\mu\rho\sigma\tau} \frac{p_\tau\, \hat t_\rho p^\lambda}{4m p^0} \frac{1}{\mcN}
  \int_{\Sigma_D} \di \Sigma \cdot p \; n_F (1 -n_F) 
   \frac{1}{T^2}u_{(\lambda}(\partial_{\sigma)} T+\C \left.\frac{\partial \mu}{\partial e}\right|_T \partial_{\sigma)}e+\C \left.\frac{\partial \mu}{\partial T}\right|_e \partial_{\sigma)}T)\,,\\
        \Delta S_\zeta &= - \epsilon^{\mu\rho\sigma\tau} \frac{p_\tau\, \hat t_\rho }{4m p^0} \frac{1}{\mcN}
  \int_{\Sigma_D} \di \Sigma \cdot p \; n_F (1 -n_F) \frac{1}{T}
  \left[\partial_\sigma T\left(\frac{1}{\mathcal{C}}+\zeta\right)+\left(\left.\frac{\partial \mu}{\partial e}\right|_T \partial_{\sigma}e+\left.\frac{\partial \mu}{\partial T}\right|_e \partial_{\sigma}T\right)\left(1+\zeta\C\right)\right]\,.
    \end{align}
\end{subequations}

From the definition \eqref{eq:variations-on-the-hyper}, we find
\begin{equation}
\left.\pdv{\mu}{T}\right|_{\Sigma_D} = \left.\pdv{\mu}{T}\right|_{e} = -\mathcal{C}^{-1}\,.
\end{equation}
Using this relation, along with $\partial \mu/\partial e=(\partial e/\partial \mu)^{-1}$, we eliminate gradients of $T$ in Eq.\ \eqref{eq:post-implicit-dS}, simplifying the expressions to
\begin{subequations}\label{eq:dS}
    \begin{align}
        \Delta S_A &=+ \epsilon^{\mu\rho\sigma\tau} \frac{p_\tau}{8m} \frac{1}{\mcN}
  \int_{\Sigma_D} \di \Sigma \cdot p \; n_F (1 -n_F) 
   \frac{1}{T^2}u_{[\rho}\partial_{\sigma]}e\left(\left.\frac{\partial e}{\partial T}\right|_\mu \right)^{-1}\, ,\\
    \Delta S_S &= \, + \epsilon^{\mu\rho\sigma\tau} \frac{p_\tau\, \hat t_\rho p^\lambda}{4m p^0} \frac{1}{\mcN}
  \int_{\Sigma_D} \di \Sigma \cdot p \; n_F (1 -n_F) 
   \frac{1}{T^2}u_{(\lambda}\partial_{\sigma)}e\left(\left.\frac{\partial e}{\partial T}\right|_\mu \right)^{-1}\,,\\
    \Delta S_\zeta &= - \epsilon^{\mu\rho\sigma\tau} \frac{p_\tau\, \hat t_\rho }{4m p^0} \frac{1}{\mcN}
  \int_{\Sigma_D} \di \Sigma \cdot p \; n_F (1 -n_F) \frac{\partial_{\sigma}e}{T}
 \left(\left.\frac{\partial e}{\partial \mu}\right|_T\right)^{-1} \,.
    \end{align}
\end{subequations}
%%%%% 

In these relations, the only nonvanishing components of the gradient of the energy density, $\partial_\mu e$, are those in the direction normal to $\Sigma_D$, as $\Sigma_D$ is defined as an iso-energy density hypersurface via Eq.\ \eqref{eq:level-surface}. 
Importantly, the integrals in Eq.\ \eqref{eq:dS} do not vanish and are expected to yield sizable differences between the two prescriptions discussed in this section. 
For instance, in the case of the isothermal freezeout hypersurface applied in high-energy heavy ion collisions, the discrepancy between the hypersurface and spacetime variations significantly affects the predicted polarization vector; with the correct prescription, the beam direction component of the polarization vector changes sign, solving the so-called local polarization sign puzzle \cite{Becattini:2021suc}. 

\subsection{Generalization to multiple chemical potentials}
We conclude this section by briefly addressing the case of multiple conserved charges, each associated with a distinct chemical potential $\mu_i$.
In this case, the variations on an iso-energy density hypersurface are related by
\begin{equation}\label{eq:multi-charge}
    \left.\frac{\partial e}{\partial T}\right|_{\mu_i}\delta T + \sum_i\left.\frac{\partial e}{\partial \mu_i}\right|_{T, \mu_{j\neq i}}\delta \mu_i = 0\,.
\end{equation}
To generalize our prescription, we identify the intensive variable $X \in \{T, \mu_i\}$ for which the magnitude \( \abs{\partial e/\partial X} \) is largest among all elements in the set, and correspondingly $\delta X$ is the smallest variation. Accordingly, the hypersurface variations $\delta Y_i$ of the remaining intensive variables $Y_i \neq X$ are approximated using their spacetime counterparts, and the variation $\delta X$ is then determined in terms of $\delta Y_i$ from Eq.\ \eqref{eq:multi-charge}.

For example, if
\begin{equation}
\left| \left(\frac{\partial e}{\partial T} \right)_{\mu_i} \right| < \left| \left( \frac{\partial e}{\partial \mu_i} \right)_{T, \mu_{j \neq i}} \right| \,,
\end{equation}
then temperature is treated as the dependent variable, and its variation is expressed as
\begin{equation}
    \delta T =- \sum_i \C_i (y-x)^\alpha \partial_\alpha \mu_i\,,\qquad \C_i = \frac{\left.\partial e/\partial \mu_i\right|_{T,\mu_{j\neq i}}}{\left.\partial e/\partial T\right|_{\mu_i}}\,,
\end{equation}
with \( \abs{\mathcal{C}_i} < 1 \) for all $i$.

If, on the other hand, the largest derivative of the energy is the one with respect to some particular $\mu_{j}$, then the hypersurface variation $\delta\mu_{j}$ is expressed in terms of those of the temperature and the other chemical potentials:
\begin{equation}
    \delta \mu_{j} =- \sum_{i\neq j} \frac{\left.\partial e/\partial \mu_i\right|_{T,\mu\neq \mu_i}}{\left.\partial e/\partial \mu_{j}\right|_{T,\mu\neq \mu_{j}}}(y-x)^\alpha \partial_\alpha \mu_i -\frac{\left.\partial e/\partial T\right|_\mu}{\left.\partial e/\partial \mu_{j}\right|_{T,\mu\neq \mu_{j}}}(y-x)^\alpha\partial_\alpha T\,.
\end{equation}

%%%%%%%%%%%%%%%%%%%%%%%%%%%%%%%%%%%%%%%%%%%%%%%%%%%%%%%
\section{Evaluation of the coefficient $\mathcal{C}$}\label{sec:evaluation}
\begin{figure}
    \centering
    \includegraphics[width=0.49\linewidth]{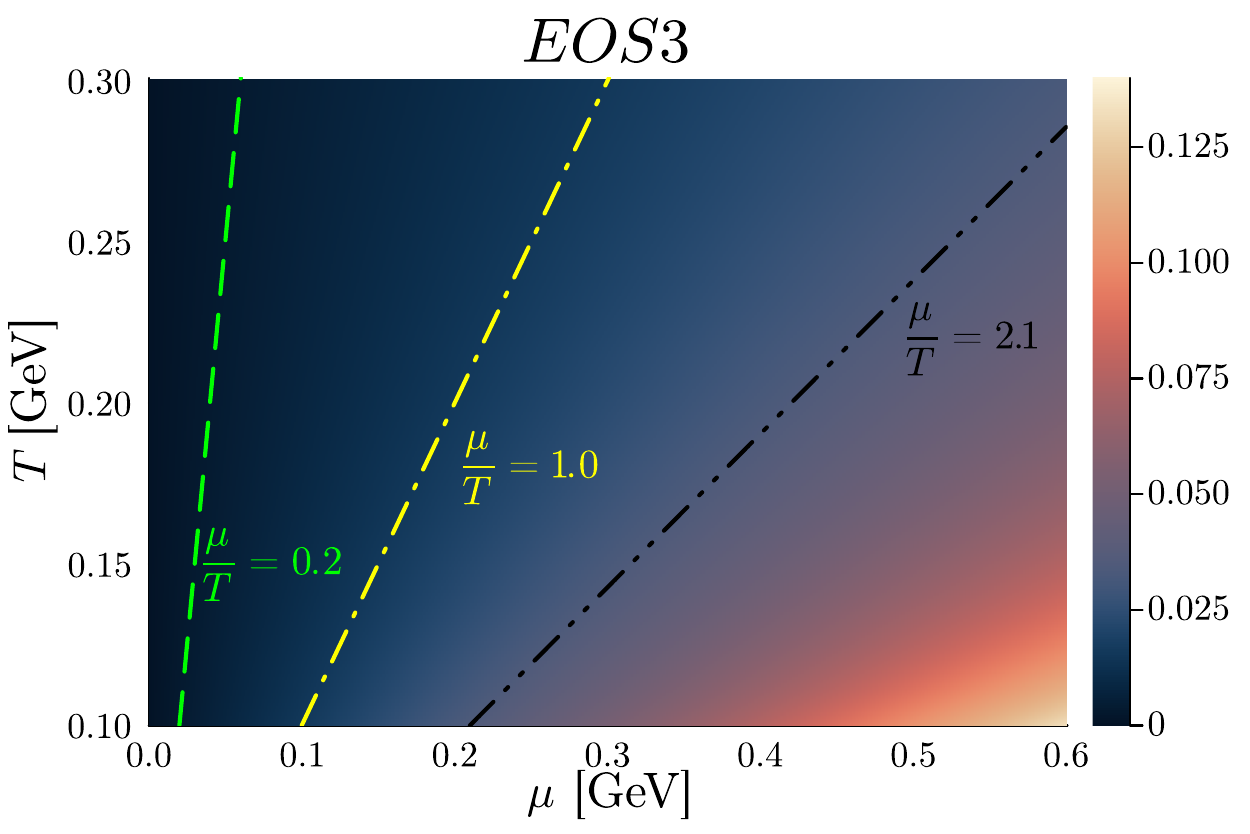}
    \includegraphics[width=0.49\linewidth]{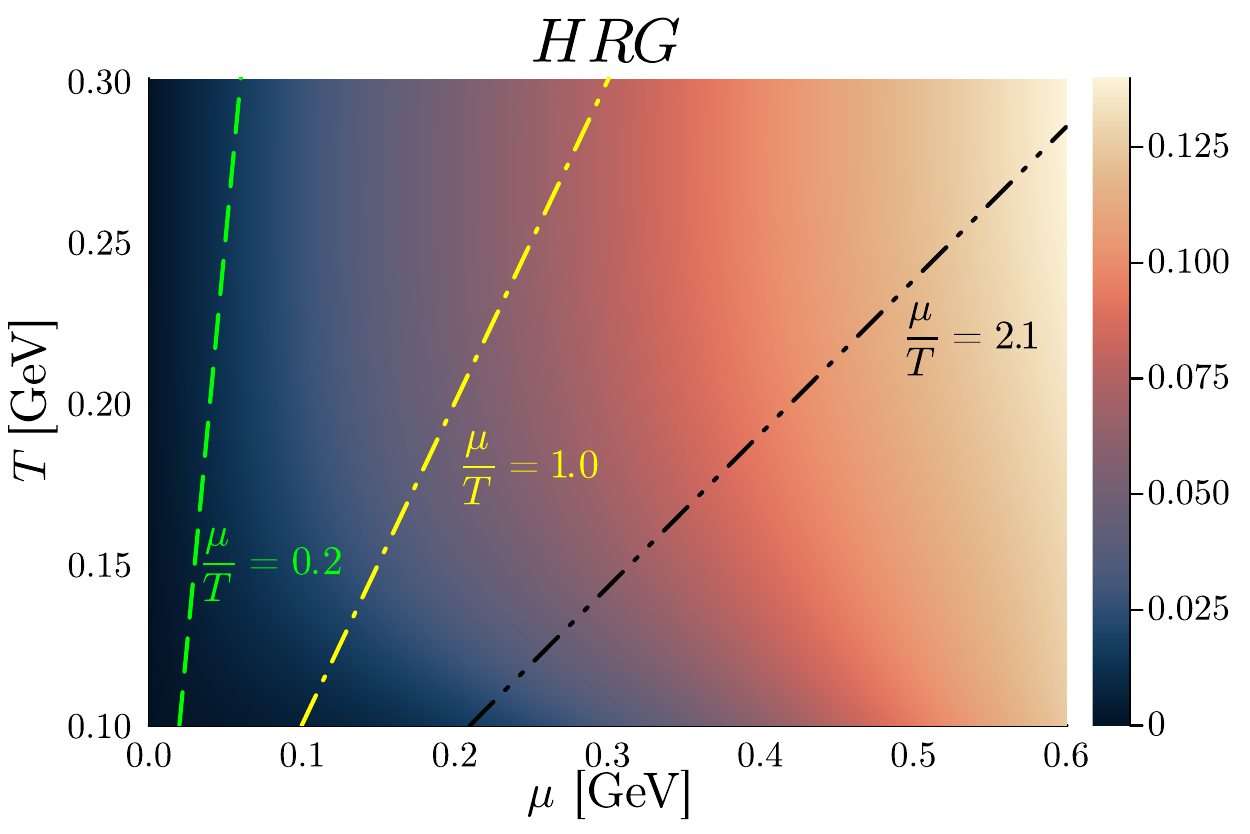}
    \caption{The values of $\mathcal{C}$ for the EOS3 equation of state (left panel) taken from Refs.\ \cite{Du:2019obx, Du:2021zqz}, reproduced in Eq.\ \eqref{eq:EOS3}, and HRG equations of state from Ref.\ \cite{Ratti2021} (right panel). Lines of constant chemical potential are also shown: Lines of constant chemical potential are also shown: green dashed for $\mu/T = 0.2$, yellow dot-dashed for $\mu/T = 1.0$, and black double dot-dashed for $\mu/T = 2.1$.}
    \label{fig:C-figure}
\end{figure}
Our results are sensitive to both the choice of equation of state and the modeling of the freezeout hypersurface. This sensitivity suggests that isothermal freezeout prescriptions should be modified in scenarios where the baryon charge density is non-negligible. The magnitude of the necessary correction is governed by the thermodynamic coefficient $\mathcal{C}$, whose value depends on the specific equation of state used.

In this section, we evaluate $\mathcal{C}$ as a function of temperature and baryochemical potential for two equations of state. 
The first, referred to as EOS3, is used in the BEShydro code \cite{Du:2019obx, Du:2021zqz}. 
It represents a generalization of the conformal equation of state for massless partons to the case of nonzero chemical potential, and is given by
\begin{equation}\label{eq:EOS3}
    e=3T^4\left[\frac{\pi^2}{90}(16+10.5 N_f)+N_f\left(\frac{1}{18}\frac{\mu^2}{T^2}+\frac{1}{324\pi^2}\frac{\mu^4}{T^4}\right)\right],
\end{equation}
where $N_f = 2.5$ is the effective number of approximately massless quark flavors (see \cite{Du:2019obx} and references therein for further details).

In addition, we compute $\mathcal{C}$ for the hadron resonance gas (HRG) equation of state (see, e.g., \cite{Ratti2021}). For this purpose, we use the HRG model implemented in the hydrodynamic code Fluid$u$m \cite{Floerchinger:2018pje,Devetak:2019lsk,Capellino:2023cxe}, which includes all hadronic resonances with vacuum masses up to 2 GeV.
The results are shown in Fig.\ \ref{fig:C-figure}. Even at large values of the baryochemical potential, we find that $\mathcal{C}$ remains small, typically $\mathcal{C} \sim 0.1 \ll 1$. This indicates that the $\mathcal{C} > 1$ regime is not realized for these equations of state. Since the chemical potential of the fireball increases as the collision energy decreases, we also show lines of constant $\mu/T$ in Fig.\ \ref{fig:C-figure}, corresponding to $\mu/T \simeq {0.2,\ 1.0,\ 2.1}$. These values are representative of the freezeout conditions at $\sqrt{s_{NN}} = {200,\ 39,\ 11.5}$ GeV, respectively, as extracted in Ref.\ \cite{STAR:2017sal}. Notably, even at the lowest beam energy available at RHIC, $\mathcal{C}$ does not exceed $10\%$.

Therefore, expressions \eqref{eq:pol-high-energy} are the relevant ones for describing polarization in heavy-ion collisions across a wide range of collision energies.
They represent geometric corrections to the isothermal freezeout prescription.
Owing to the smallness of $\mathcal{C}$, these corrections are expected to be modest in magnitude; however, they may become significant when computing local polarization in low-energy heavy-ion collisions.
Furthermore, different equations of state used in hydrodynamic simulations, particularly those featuring a critical point, may enhance the relevance of the effects discussed here.

\section{Conclusions}\label{sec:conclusions}
In this work, we developed a geometric-thermodynamic framework that generalizes the isothermal freezeout prescription used in Ref.\ \cite{Becattini:2021iol} to describe local polarization in heavy-ion collisions.  
Our approach is based on the assumption that freezeout occurs on an iso-energy density hypersurface, and it naturally reduces to the isothermal case when the energy density depends only on temperature.  
We derived a set of polarization formulae that account for geometric corrections to the standard prescription, characterized by the thermodynamic coefficient $\mathcal{C}$.  

By evaluating $\mathcal{C}$ for two different equations of state, we found that it remains small---typically below 10\%---even at low collision energies.  
This suggests that the isothermal freezeout approximation is robust across a broad range of energies.  
Nonetheless, the corrections introduced by our generalized prescription may still play a non-negligible role in refining predictions of local polarization observables, particularly in low-energy collisions or in scenarios involving a more complex equation of state, such as one featuring a critical point.

\section*{Acknowledgements}
We thank A.\ Mazeliauskas for suggesting this problem and participating in the first stages of its study. 
A.\ P.\ is supported by the U.S. Department of Energy under Grant DE-FG88ER40388.
M.\ S.\ is supported by the Deutsche Forschungsgemeinschaft (DFG, German Research Foundation) through the Collaborative Research Center CRC-TR 211 ``Strong-interaction matter under extreme conditions''- project number 315477589 - TRR 211.

\appendix
\section{Value of $\mathcal{C}$ in terms of other thermodynamic coefficients and functions}\label{app:coefficient}
The coefficient $\mathcal{C}$ introduced in the text has been defined as:
\begin{equation}
    \mathcal{C}=\frac{\left.\partial e/\partial \mu\right|_{T}}{\left.\partial e /\partial T\right|_\mu}.
\end{equation}
The derivatives involved in this ratio can be expressed in terms of other thermodynamic coefficients. 
In particular, we express $\mathcal{C}$ in terms of the isothermal compressibility $\kappa_T$, the expansion coefficient $\alpha$, and heat capacities at constant volume and pressure, $c_V$ and $c_P$.
The definitions of these coefficients are (see, for instance, \cite{Floerchinger:2015efa}):
\begin{subequations}\label{eq:thermodynamic-coefficients}
\begin{align}
    \kappa_T&=\frac{1}{n}\left(\frac{\de n}{\de P}\right)_T,
    \qquad \qquad
    \alpha = -\frac{1}{n}\left(\frac{\de n}{\de T}\right)_P,\\
    c_V &= T\left(\frac{\de s}{\de T}\right)_n,
    \qquad \qquad
    c_P = \frac{Tn}{s}\left(\frac{\de s/n}{\de T}\right)_P=\frac{T}{s}\left(\frac{\de s}{\de T}\right)_P+T\alpha.
\end{align}    
\end{subequations}

We now recall several thermodynamic identities: the Euler relation, which follows from extensivity; the first law of thermodynamics; and the Gibbs-Duhem relation. 
These read:
\begin{subequations}
\begin{align}\label{eq:euler}
    e+P&=Ts+\mu n\,,\\
    \label{eq:first-law}
     \di e &= T\di s+\mu \di n\,,\\
     \label{eq:Gibbs-Duhem}
     \di P &= s\di T+n \di \mu\,.
\end{align}
\end{subequations}

Assuming that $s$ and $n$ are functions of $T$ and $\mu$, we can express Eq.\ \eqref{eq:first-law} as
\begin{equation}
    \di e = \left[\left(T\frac{\de s}{\de T}+\mu \frac{\de n}{\de T}\right)\right]_\mu \di T + \left[\left(T\frac{\de s}{\de \mu}+\mu \frac{\de n}{\de \mu}\right)\right]_T \di \mu\,,
\end{equation}
which allows the identification of $(\de e/\de T)_\mu$ and $(\de e/\de \mu)_T$ as
\begin{equation}\label{eq:e-ders}
    \left(\pdv{e}{T}\right)_\mu = T\left(\frac{\de s}{\de T}\right)_\mu+\mu \left(\frac{\de n}{\de T}\right)_\mu\,,\qquad  \left(\pdv{e}{\mu}\right)_T = T\left(\frac{\de s}{\de \mu}\right)_T+\mu \left(\frac{\de n}{\de \mu}\right)_T\,.
\end{equation}
To evaluate these derivatives, we first use Jacobian manipulations (see, e.g., Ref.\ \cite{Landau_vol5}) to write:
\begin{equation}
    \left(\frac{\de n}{\de T}\right)_\mu = \frac{\partial(n,\mu)}{\partial(T,\mu)} = \frac{\partial(n,\mu)}{\partial(T,P)}\frac{\partial(T,P)}{\partial(T,\mu)}\,,\qquad
    \left(\frac{\de n}{\de \mu}\right)_T = \frac{\partial(n,T)}{\partial(\mu,T)} = \frac{\partial(n,T)}{\partial(P,T)}\frac{\partial(P,T)}{\partial(\mu,T)}\,, 
\end{equation}
and then use definitions \eqref{eq:thermodynamic-coefficients} as well as Eq.\ \eqref{eq:Gibbs-Duhem} to arrive at
\begin{equation}\label{eq:n-ders}
    \left(\frac{\de n}{\de T}\right)_\mu=-\alpha n+\kappa_Ts\,n\,, 
    \qquad \qquad
    \left(\frac{ \de n}{\de \mu}\right)_T=\kappa_T n^2\,.
\end{equation}
We now evaluate the derivatives of the entropy density. 
To proceed, we first write
\begin{equation}\label{eq:d-s}
    \dd{s} = \left(\frac{ \de s}{\de T}\right)_\mu \di T + \left(\frac{ \de s}{\de \mu}\right)_T \di \mu\,.
\end{equation}
Using this identity at constant $n$ and $P$, we arrive at the following system of equations:
\begin{equation}
    \left(\pdv{s}{T}\right)_n=\left(\frac{\de s}{\de T}\right)_\mu +\left(\pdv{s}{\mu}\right)_T \left(\pdv{\mu}{T}\right)_n\,, 
    \qquad \qquad
   \left(\pdv{s}{T}\right)_P=\left(\frac{\de s}{\de T}\right)_\mu +\left(\pdv{s}{\mu}\right)_T \left(\pdv{\mu}{T}\right)_P\,,
\end{equation}
where the left-hand sides of both equations are given by definitions \eqref{eq:thermodynamic-coefficients}.
Solving the system of equations for our desired derivatives yields
\begin{align}\label{eq:s-ders-1}
    \left(\pdv{s}{\mu}\right)_T &=   \left[\left(\pdv{s}{T}\right)_n-\left(\pdv{s}{T}\right)_P\right]\left[\left(\pdv{\mu}{T}\right)_n+\frac{s}{n}\right]^{-1}\,,\\
    \left(\pdv{s}{T}\right)_\mu &=   \left[\left(\pdv{\mu}{T}\right)_n\left(\pdv{s}{T}\right)_P+\frac{s}{n}\left(\pdv{s}{T}\right)_n\right]\left[\left(\pdv{\mu}{T}\right)_n+\frac{s}{n}\right]^{-1}\,,
\end{align}
where we have used Eq.\ \eqref{eq:Gibbs-Duhem} to write
\begin{equation}
    \left(\pdv{\mu}{T}\right)_P = -\frac{s}{n}\,.
\end{equation}
We then use Maxwell relation and Eq.\ \eqref{eq:n-ders} to obtain
\begin{equation}
    \left(\pdv{\mu}{T}\right)_n = - \left(\pdv{n}{T}\right)_\mu \left(\pdv{\mu}{n}\right)_T =  \frac{\alpha-\kappa_T s}{n\kappa_T}\;.
\end{equation}
Employing this relation together with the definitions \eqref{eq:thermodynamic-coefficients} in Eq.\ \eqref{eq:s-ders-1}, we find
\begin{align}\label{eq:s-ders}
    \left(\frac{\de s}{\de T}\right)_\mu &=\frac{s [c_V \kappa_T +(\alpha T -c_P)(\kappa_T  s-\alpha)]}{\alpha  T},\\
   \left(\frac{\de s}{\de \mu}\right)_T &= \frac{\kappa_T  n (-c_P s+c_V+\alpha  s T)}{\alpha  T}.
\end{align}
Finally, using the above equations and Eq.\ \eqref{eq:n-ders} in Eq.\ \eqref{eq:e-ders}, and inserting the result into Eq.\ \eqref{eq:def-C}, we arrive at 
\begin{equation}
    \mathcal{C}
   =\frac{n\,\kappa_T  (c_V-s\,c_P+\alpha h)}{(s\,\kappa_T -\alpha)(c_V-s\,c_P+\alpha h)-\alpha c_V}\,,
\end{equation}
where $h\equiv e + P$ is the enthalpy density.

%%%%%%%%%%%%%%%%%%%%%%%%%%%%%%%%%%%%%%%%%%%%%%%%%%%%%%%%%%%%%%%%%%%%%%%%%%%%%%%%%%%%%%%%%%%%%%%%%%%%%
%                                       BIBLIOGRAPHY
%%%%%%%%%%%%%%%%%%%%%%%%%%%%%%%%%%%%%%%%%%%%%%%%%%%%%%%%%%%%%%%%%%%%%%%%%%%%%%%%%%%%%%%%%%%%%%%%%%%%%%
% \bibliographystyle{ieeetr}
\bibliography{biblio}

\end{document}

%% file: commands.tex
\def\!{\mskip-\thinmuskip}

\newcommand{\di}{{\rm d}}

\newcommand{\wJ}{\widehat{J}}
\newcommand{\wT}{\widehat{T}}
\newcommand{\wj}{\widehat{j}}
\newcommand{\wQ}{\widehat{Q}}
\newcommand{\wP}{\widehat{P}}

\newcommand{\wrho}{\widehat{\rho}}

\newcommand{\tr}{{\rm tr}}

\newcommand{\de}{\partial}
\newcommand{\C}{\mathcal{C}}
\newcommand{\mcN}{\mathcal{N}}